\documentclass[aps,prd,superscriptaddress,amsmath,twocolumn,nofootinbib,10pt]{revtex4}

\usepackage{color,hangcaption,hhline,psfrag,rotating,amssymb}
\usepackage[hang,nooneline]{subfigure}
\usepackage{dcolumn}
\usepackage{verbatim}
\usepackage{bm}

\begin{document}
\title{Strange and charm quark contributions to the anomalous magnetic moment of the muon}

\author{Bipasha Chakraborty}
\affiliation{SUPA, School of Physics and Astronomy, University of Glasgow, Glasgow, G12 8QQ, UK}
\author{C.~T.~H.~Davies}
\email[]{christine.davies@glasgow.ac.uk}
\affiliation{SUPA, School of Physics and Astronomy, University of Glasgow, Glasgow, G12 8QQ, UK}
\author{G.~C.~Donald}
\affiliation{Institut f\"{u}r Theoretische Physik, Universit\"{a}t Regensburg, 93040 Regensburg, Germany}
\author{R.~J.~Dowdall}
\affiliation{DAMTP, University of Cambridge, Wilberforce Road, Cambridge, CB3 0WA, UK}
\author{J.~Koponen}
\affiliation{SUPA, School of Physics and Astronomy, University of Glasgow, Glasgow, G12 8QQ, UK}
\author{G. P. Lepage}
\affiliation{Laboratory for Elementary-Particle Physics, Cornell University, Ithaca, New York 14853, USA}
\affiliation{DAMTP, University of Cambridge, Wilberforce Road, Cambridge, CB3 0WA, UK}
\collaboration{HPQCD collaboration}
\homepage{http://www.physics.gla.ac.uk/HPQCD}
\noaffiliation
\author{T. Teubner}
\affiliation{Department of Mathematical Sciences, University of Liverpool, Liverpool, L69 3BX, UK}

\date{\today}

\begin{abstract}
We describe a new technique to determine the contribution to the anomalous magnetic moment of the muon coming from the hadronic vacuum polarization using lattice QCD. 
Our method reconstructs the Adler function, using Pad\'{e} approximants, 
from its derivatives at $q^2=0$ 
obtained simply and accurately from time-moments of the vector 
current-current correlator at zero spatial momentum. 
We test the method using strange quark correlators 
on large-volume gluon field configurations that include the effect of 
up and down (at physical masses), strange and charm quarks in 
the sea 
at multiple values of the lattice spacing and multiple volumes and show that 1\% accuracy is achievable.  
For the charm quark contributions we use our previously determined 
moments with up, down and strange quarks in the sea on very fine lattices.  
We find the (connected) contribution to the anomalous moment 
from the strange quark vacuum polarization to be 
$a_\mu^s = 53.41(59) \times 10^{-10}$, and from 
charm to be $a_\mu^c = 14.42(39)\times 10^{-10}$. 
These are in good agreement with flavour-separated results from 
non-lattice methods, given caveats about the comparison. 
The extension of our method to the light quark contribution and to that 
from the quark-line disconnected diagram is straightforward. 
\end{abstract}


\maketitle

\section{Introduction} 
\label{sec:intro}

\begin{figure}
\centering
\includegraphics[width=0.25\textwidth]{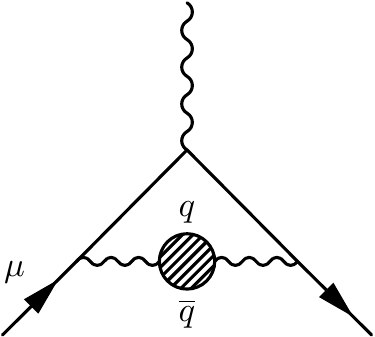}
\caption{
The hadronic vacuum polarization contribution to the 
muon anomalous magnetic moment is represented as a shaded 
blob inserted into the photon propagator (represented 
by a wavy line) that corrects the point-like 
photon-muon coupling at the top of the diagram. 
}
\label{fig:hvp}
\end{figure}

The magnetic moment of the muon can be determined extremely accurately 
in experiment. Its anomaly, defined as the fractional difference 
of its gyromagnetic ratio from the naive value of 2 ($a_{\mu}=(g-2)/2$) is known 
to 0.5 ppm~\cite{Bennett:2006fi}. 
The anomaly arises 
from muon interactions with a cloud of virtual 
particles. However, the theoretical calculation of $a_{\mu}$ in the Standard 
Model shows a discrepancy with the experimental result of 
about $25(9) \times 10^{-10}$~\cite{Aoyama:2012wk, Hagiwara:2011af, Davier} 
which could be an exciting indication of the existence of new 
virtual particles. Improvements of a factor of 4 in the experimental uncertainty 
are expected and improvements in the theoretical determination 
would make the discrepancy (if it remains) really compelling~\cite{snowmass}. 

The current theoretical uncertainty is dominated by that from 
the hadronic vacuum polarization (HVP) contribution, depicted in Fig.~\ref{fig:hvp}.   
This contribution is currently determined most accurately from 
experimental results on $e^+e^- \rightarrow$ hadrons 
or from $\tau$ decay to be of size $700 \times 10^{-10}$ 
with a 1\% error~\cite{Hagiwara:2011af, Davier}. 
Higher order contributions from QCD processes, such as the 
hadronic light-by-light diagram, have larger percentage uncertainty 
but make an order of magnitude smaller contribution, so 
do not contribute as much to the overall theoretical 
uncertainty. 

In 2002 Blum~\cite{Blum:2002ii} showed how to express 
the HVP contribution in terms of the vacuum polarization function evaluated 
at Euclidean~$q^2$, which greatly facilitates its calculation from lattice QCD. 
He reviews the status of such calculations in~\cite{Blum:2013qu},
which now include sea quarks~\cite{aubin-blum, boyle, renner, wittig}. 
Lattice QCD calculations, however, are not yet 
at a level where they can compete with the precision 
of values that use experimental results. 
A key problem has been that of determining 
the hadron vacuum polarization at small 
values of the (Euclidean) squared 4-momentum, $q^2$, of 
$\mathcal{O}(m_{\mu}^2)$ which are the 
key ones contributing to $a_{\mu}$. 
Extrapolating from higher values of $q^2$
leads to model uncertainties and direct calculation 
at lower $q^2$ using `twisted boundary conditions' 
produces noisy results. Efforts are underway to 
improve both of these approaches~\cite{aubin-twisted, Aubin:2012me}. 
See also~\cite{Feng:2013xsa, Francis:2013fzp, Gregory:2013taa}. 

Here we sidestep this issue by expressing the $g-2$
HVP contribution in terms of
a small number of derivatives of the hadronic vacuum polarization 
function evaluated at $q^2=0$. In effect, we work upwards from 
$q^2=0$ into the region of important, but still very small, $q^2$ 
values. 
The advantage of this method is that the derivatives are readily and 
accurately 
computed from time-moments of the current-current correlator
at zero spatial momentum.

We can approximate the hadronic vacuum polarization function by its Taylor 
expansion when $q^2$ is of order $m_\mu^2$, but the series diverges
when $q$ is of order or larger than 
the threshold energy for real hadron production ($2m_\pi$ for $u/d$ quarks). Contributions
from high momenta, say~$q\ge1$\,GeV, are suppressed by $(m_\mu/q)^2$ but 
remain important if one desires better than 1\%~precision. To deal with high
momenta, we replace the Taylor expansion by its Pad\'{e} 
approximants~\cite{Pade}. 
Successive orders of Pad\'{e} approximant converge 
to the exact vacuum polarization function for
all positive (Euclidean)~$q^2$~\cite{Baker:1969, Baker:1996}. 
This follows from the dispersion relation for the vacuum 
polarization~\cite{Aubin:2012me}. 
As we will show, only a few orders are needed to achieve 1\%~accuracy
or better. The Pad\'e approximants 
capture the entire contribution for all~$q^2$, through analytic 
continuation from low~$q^2$ to high~$q^2$,  
and so, unlike in some other approaches to HVP,
we need not calculate high-$q^2$ contributions 
using perturbation theory 
(since this would constitute double-counting). 

A further advantage of our approach is that it works with both
local lattice approximations to the vector current, and 
exactly conserved but nonlocal approximations. Local approximations
are easy to code 
and less noisy than nonlocal approximations, and so are widely used
in lattice simulations. The fact that they are not exactly 
conserved
leads to nonzero contributions 
to the vacuum polarization function $\Pi^{\mu\nu}(q^2)$ at $q^2=0$, 
but such contributions are discarded automatically when we express 
$g-2$ in terms of derivatives of $\Pi^{\mu\nu}$.

In this paper we illustrate our method by applying it to correlators made of 
$s$ quarks, using well-tuned $s$ quark 
masses on gluon field configurations that 
include up, down, strange and charm quarks in 
the sea. The sea up and down quarks have physical 
values, so no chiral extrapolation to the physical 
point is needed. We have three values of the 
lattice spacing, allowing good control of 
the extrapolation to zero lattice spacing. 
A study on three different volumes at one 
value of the lattice spacing allows us also 
to control finite volume effects. 

We also give a result for the much smaller charm contribution, using 
moments determined previously by us~\cite{Donald:2012ga,jpsi} on configurations covering 
a large range of lattice spacing values and 
including up, down and strange quarks in the sea. 

The next section gives 
details of the lattice calculation and tests of our approach; 
we then discuss our results and give our conclusions. 

\section{Lattice Calculation}
\label{sec:lattice}

\begin{table*}
\caption{
The lattice QCD gluon field configurations used here come from the MILC
collaboration~\cite{Bazavov:2010ru, Bazavov:2012uw}. 
$\beta=10/g^2$ is the QCD gauge coupling, and $w_0/a$~\cite{fkpi} 
gives the lattice spacing, $a$, in terms of the Wilson flow parameter, 
$w_0$~\cite{Borsanyi:2012zs}. 
We take $w_0$=0.1715(9)\,fm fixed from $f_{\pi}$~\cite{fkpi}. $L$ and $T$ 
give the length in the space and time directions for each lattice. 
$am_\ell^\mathrm{sea}, am_s^\mathrm{sea}$ 
and $am_c^\mathrm{sea}$ are the light ($m_\ell\equiv m_u=m_d$), 
strange, and charm sea quark masses in lattice units and
$am_s^\mathrm{val}$, the valence strange quark mass, tuned from 
the mass of the $\eta_s$, $aM_{\eta_s}$. $Z_{V,\overline{s}s}$ gives the vector 
current renormalization factor obtained nonperturbatively~\cite{Chakraborty:2014zma}.
The lattice spacings are approximately 0.15\,fm for sets~1--2, 
0.12\,fm for sets~3--8, and 0.09\,fm for sets~9--10. 
Light sea-quark masses range from $m_s/5$ to the physical value 
and lattice volumes ranging from~2.5\,fm to~5.8\,fm.
The number of 
configurations is given 
in the final column. We used 16 time sources on each (12 on sets 1 and 2).   
}
\label{tab:params}
\begin{ruledtabular}
\begin{tabular}{llllllllllll}
Set & $\beta$ &  $w_0/a$ & $am_\ell^\mathrm{sea}$ & $am_s^\mathrm{sea}$ 
& $am_c^\mathrm{sea}$ 
& $am_s^\mathrm{val}$ & $aM_{\eta_s}$ 
& $Z_{V,\overline{s}s}$ & $L/a\times T/a$ & $n_{\mathrm{cfg}}$ \\
\hline
1  & 5.80 & 1.1119(10)  & 0.01300 & 0.0650 & 0.838 & 0.0705 & 0.54024(15)  & 0.9887(20) & 16$\times$48 & 1020 \\
2  & 5.80 & 1.13670(50) & 0.00235 & 0.0647 & 0.831 & 0.0678 & 0.526799(81) & 0.9887(20) & 32$\times$48 & 1000 \\
\hline
3  & 6.00 & 1.3826(11)  & 0.01020 & 0.0509 & 0.635 & 0.0541 & 0.43138(12)  & 0.9938(17) & 24$\times$64 & 526 \\
4  & 6.00 & 1.4029(9) & 0.00507 & 0.0507 & 0.628 & 0.0533 & 0.42664(9) & 0.9938(17) & 24$\times$64 & 1019  \\
5  & 6.00 & 1.4029(9) & 0.00507 & 0.0507 & 0.628 & 0.0533 & 0.42637(6) & 0.9938(17) & 32$\times$64 & 988  \\
6  & 6.00 & 1.4029(9) & 0.00507 & 0.0507 & 0.628 & 0.0507 & 0.41572(14)  & 0.9938(17) & 32$\times$64 & 300  \\
7  & 6.00 & 1.4029(9) & 0.00507 & 0.0507 & 0.628 & 0.0533 & 0.42617(9) & 0.9938(17) & 40$\times$64 & 313  \\
8  & 6.00 & 1.4149(6) & 0.00184 & 0.0507 & 0.628 & 0.0527 & 0.423099(34) & 0.9938(17) & 48$\times$64 & 1000 \\
\hline
9  & 6.30 & 1.8869(39)  & 0.00740 & 0.0370 & 0.440 & 0.0376 & 0.31384(9) & 0.9944(10) & 32$\times$96 & 504 \\
10 & 6.30 & 1.9525(20)  & 0.00120 & 0.0363 & 0.432 & 0.0360 & 0.30480(4) & 0.9944(10) & 64$\times$96 & 621  \\
\end{tabular}
\end{ruledtabular}
\end{table*}

For the strange quark contribution to $a_{\mu}$ we use the 
Highly Improved Staggered Quark (HISQ) action~\cite{HISQ_PRD} for all quarks. 
This has small discretization errors~\cite{HISQ_PRD, HISQ_PRL, Dsdecayconst} 
and is numerically very fast. We calculate HISQ $s$ quark propagators on 
gluon field configurations generated by the MILC collaboration that 
include $u$, $d$, $s$ and $c$ quarks in the sea using 
the HISQ formalism~\cite{Bazavov:2010ru, Bazavov:2012uw}. 
Details of the ensembles are given in Table~\ref{tab:params}. 
They range in lattice spacing from 0.15\,fm down to 0.09\,fm with the spatial length 
of the lattice as large as 5.6\,fm on the finest lattices.  
At each lattice spacing we have two values of the average $u/d$ quark mass: one fifth 
the $s$ quark mass and the physical value ($m_s/27.5$~\cite{durrmsml}). 
The tuning of the valence masses is more critical than that of 
the sea, so the valence and sea $s$ masses differ slightly. 
We tune the valence $s$ mass accurately using 
the mass of the $\eta_s$, a pseudoscalar 
pure $s\overline{s}$ meson which does not occur in the real world. 
In lattice QCD, where the $\eta_s$ can be prevented from mixing with other mesons, 
its properties can be very accurately determined~\cite{Davies:2009tsa}. 
Its mass (688.5(2.2)\,MeV~\cite{fkpi}) is very sensitive to the $s$ quark 
mass, making it useful for tuning. 
At a third value of the $u/d$ quark mass, one tenth of the $s$ quark mass, 
we have three different volumes to test for finite volume effects. These 
are sets 4, 5 and 7 and correspond to a lattice length in units of 
the $\pi$ meson mass~\cite{fkpi} of $M_{\pi}L=$ 3.2, 4.3 and 5.4. In addition we 
de-tuned the valence $s$ quark mass there by 5\% (set 6) to test for tuning effects.  

The $s$ quark propagators are combined into a correlator with a local vector 
current at either end. The end point is summed over spatial sites on a timeslice 
to set the spatial momentum to zero. The source is created from a set of 
U(1) random numbers over a timeslice for improved statistics. 
The local current is not the conserved vector current 
for this quark action and must be normalised. We do this completely 
nonperturbatively by demanding that the vector form factor for this current be 1 between two 
equal mass mesons at rest ($q^2=0$)~\cite{Chakraborty:2014zma}. 
The $Z$ factors are given in Table~\ref{tab:params}. They differ from~1 
by at most 1\% (on the 0.15\,fm lattices) and vary from one lattice 
spacing to another by less than 0.5\%. We therefore only calculate them for 
the $m_l/m_s$=0.2 ensembles at each lattice spacing.   
At large time separations between source and sink these correlators give 
the mass and decay constant of the $\phi$ meson~\cite{Chakraborty:2014zma}. 
Here we are concerned with the properties of the correlation function at the shorter times 
that feed into the theoretical determination of~$a_{{\mu},\mathrm{HVP}}$.  

The contribution to the muon anomalous magnetic moment 
from the HVP associated with a given quark flavour, $\mathrm{f}$, is obtained 
by inserting the quark vacuum polarization into the photon propagator~\cite{Blum:2002ii}: 
\begin{equation}
a_{\mu, \mathrm{HVP}}^{(\mathrm{f})} = \frac{\alpha}{\pi} \int_0^{\infty} dq^2 f(q^2) (4\pi\alpha Q_{\mathrm{f}}^2) \hat{\Pi}_{\mathrm{f}}(q^2)  
\label{eq:amu}
\end{equation}
where $\alpha \equiv \alpha_{\mathrm{QED}}$ and $Q_{\mathrm{f}}$ is the electric charge of quark $\mathrm{f}$ in units 
of $e$. Here 
\begin{equation}
f(q^2) \equiv \frac{m_{\mu}^2 q^2 A^3 (1-q^2A)}{1+m_{\mu}^2q^2A^2} 
\label{eq:f}
\end{equation}
where
\begin{equation}
A \equiv \frac{\sqrt{q^4 + 4 m_{\mu}^2q^2} - q^2}{2m_{\mu}^2q^2}. 
\label{eq:A}
\end{equation}
Note that in our calculation we have ignored 
quark-line-disconnected contributions to the HVP. These 
are suppressed by quark mass factors since they would vanish 
for equal mass $u$, $d$ and $s$ quarks since 
$\sum_{u,d,s} Q_{\mathrm{f}} = 0$~\cite{Blum:2002ii}. 

The quark polarization tensor is the Fourier transform 
of the vector current-current correlator. 
For spatial currents at zero spatial momentum 
\begin{equation}
\Pi^{ii}(q^2) =
q^2{\Pi}(q^2) = a^4 \sum_t e^{iqt} \sum_{\vec{x}}\langle j^{i}(\vec{x},t)j^{i}(0) \rangle 
\label{eq:pi}
\end{equation}
with $q$ the Euclidean energy. We need the renormalized 
vacuum polarization function, $\hat\Pi(q^2)\equiv\Pi(q^2)-\Pi(0)$. 
Time-moments of the correlator give the derivatives 
at $q^2=0$ of $\hat{\Pi}$~\cite{disc}
(see, for example, \cite{Allison:2008xk,McNeile:2010ji}): 
\begin{eqnarray}
G_{2n} &\equiv& a^4 
\sum_t \sum_{\vec{x}} t^{2n} Z_V^2 \langle j^{i}(\vec{x},t)j^{i}(0) \rangle  \nonumber \\
&=& (-1)^n \left. \frac{\partial^{2n}}{\partial q^{2n}} q^2\hat{\Pi}(q^2) \right|_{q^2=0} .
\label{eq:G}
\end{eqnarray}
Here we have allowed for a renormalization factor $Z_V$ for the lattice vector 
current. Note that  time-moments remove any contact terms between the two 
currents\footnote{The vector current need not be exactly 
conserved, provided that it is renormalized correctly with $Z_V$ because: a) there are no contributions from contact terms in the moments, and b) the only lattice operators that can mix with the vector current have higher dimension and so are suppressed by powers of~$a^2$.}. $G_{2n}$ is easily calculated from the correlators 
calculated in lattice QCD, remembering that time runs from 0 at the origin in both 
positive and negative directions to a maximum value of  
$T/2$ in the centre of the lattice. 

Defining 
\begin{equation}
\hat{\Pi}(q^2) = \sum_{j=1}^{\infty} q^{2j} \Pi_j  
\label{eq:pihat}
\end{equation}
then 
\begin{equation}
\Pi_j = (-1)^{j+1} \frac{G_{2j+2}}{(2j+2)!} \, .
\label{eq:derivs}
\end{equation}
To evaluate the contribution to $a_{\mu}$ we will 
replace $\hat{\Pi}(q^2)$ with its $[n,n]$ and $[n,n-1]$ Pad\'{e} approximants
derived from the~$\Pi_j$~\cite{Pade}.
We perform the $q^2$ integral numerically. 

Eq.~(\ref{eq:G}) is, of course, approximate when the the temporal extent~$T$
of the lattice is finite. 
This error is exponentially suppressed, and usually
negligible, because $G(t)$ falls to zero quickly with increasing $|t|$
($\le T/2$) and has effectively vanished well before $|t|$ gets to edge of 
the lattice at~$T/2$. At large $|t|$ 
the correlator is dominated by the lowest-energy 
vector state in the simulation\,---
\begin{equation}
    G(t) \to a_0 \left( \mathrm{e}^{-E_0|t|} + \mathrm{e}^{-E_0(T-|t|)}\right)
\end{equation}
---\, so that terms containing $T$ are suppressed by a factor of
$\exp(-E_0T/2)$. Such terms become important for high order moments, 
since $t^nG(t)$ peaks at $t\approx n/E_0$ for large $n$, but they are
negligible for the moments of interest here. Their impact on 
$a_{\mu}^s$ is easily estimated (see Section~\ref{sec:discussion}): 
for example, they enter at the level of 0.002\% in our analysis for 
configuration set~10 in Table~\ref{tab:params}. 

\begin{figure}
\centering
\includegraphics[width=0.475\textwidth]{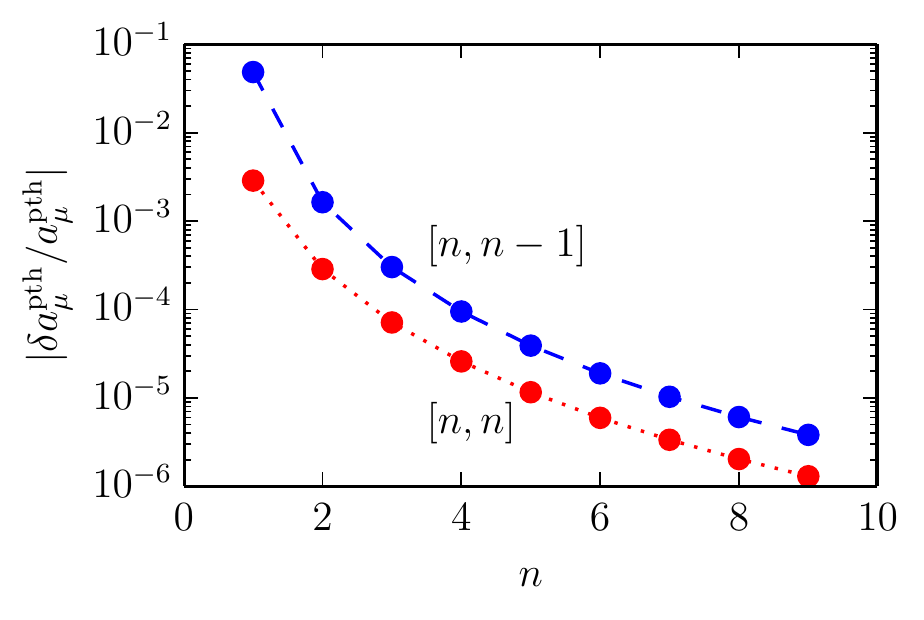}
\caption{Fractional error in the muon anomaly~$a_\mu$ 
caused by replacing the 
quark vacuum polarization from one-loop perturbation theory
with its $[n,n]$ and $[n,n-1]$ Pad\'{e} approximants. 
The exact result is always between the $[n,n-1]$ and
$[n,n]$ approximants. The quark mass is set equal to the
kaon mass in this test case.
}
\label{fig:test}
\end{figure}
\begin{figure}
\centering
\includegraphics[width=0.475\textwidth]{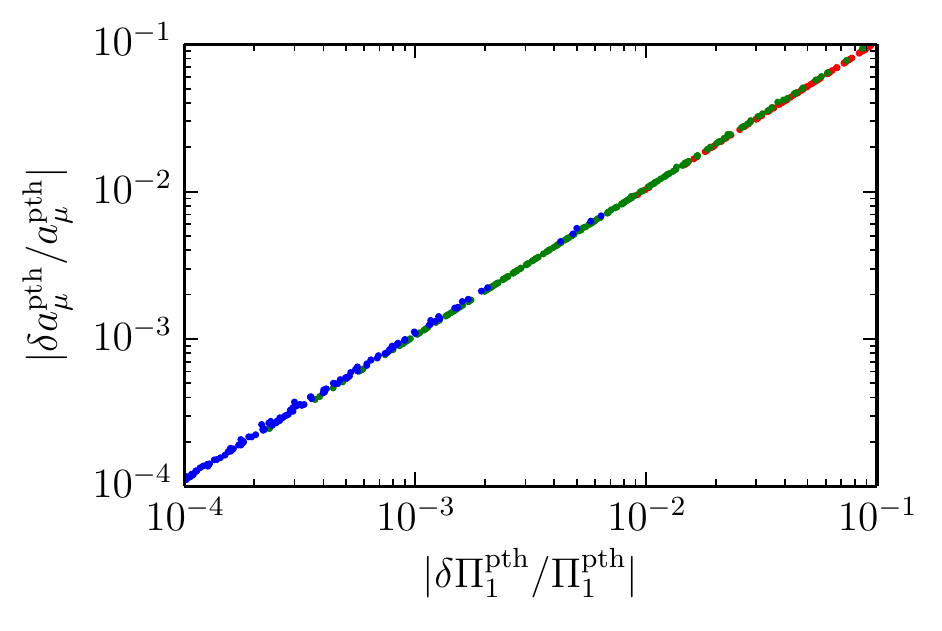}
\caption{Fractional errors in the muon anomaly~$a_\mu$
caused by adding random noise
to the moments of the one-loop vacuum polarization. Results
are shown for 400~different simulations, each with different
amounts of random noise. They are plotted against the fractional
uncertainty in the leading moment. In each simulation, $[n,n]$~Pad\'{e}
approximants are used where $n$ is increased until
results for $a_\mu^\mathrm{pth}$ converge or spurious poles appear in the
approximant. Color indicates the value of $n$ used: red for $[1,1]$,
green for $[2,2]$, and blue for $[3,3]$. Simulation results
agreed with the exact result to within $\pm1\sigma$ in~70\% of the 
simulations, as expected. The quark mass is set equal to the 
kaon mass in each case.
}
\label{fig:test2}
\end{figure}

\newcommand{\cPi}[1]{\hfill $\Pi_#1$ \hfill\mbox{}}
\begin{table*}
    \caption{Columns 2-5 give the Taylor coefficients $\Pi_j$ (Eq.~\ref{eq:pihat}),
    in units of $1/\mathrm{GeV}^{2j}$,
    for each of the lattice data sets in Table~\ref{tab:params}.
The errors given include statistics and the (correlated) uncertainty from setting the 
lattice spacing using $w_0$, which dominates. 
    Estimates of the connected contribution from $s$-quarks to $a_{\mu,\mathrm{HVP}}$ are given for each
    of the $[1,0]$, $[1,1]$, $[2,1]$ and $[2,2]$ Pad\'e approximants in 
columns 6-9;
    results are multiplied by~$10^{10}$.}
    \label{tab:s-data}
    \begin{ruledtabular}
        \begin{tabular}{ccccccccc}
            Set &  \cPi1 & \cPi2 & \cPi3 & \cPi4 
            & $[1,0]\times10^{10}$ & $[1,1]\times10^{10}$ 
            & $[2,1]\times10^{10}$ & $[2,2]\times10^{10}$
             \\ \hline
\hline
1 & $0.06598(76)$ & $-0.0516(11)$ & $0.0450(15)$ & $-0.0403(19)$ & 58.11(67) & 53.80(59) & 53.95(59) & 53.90(59)\\
2 & $0.06648(75)$ & $-0.0523(11)$ & $0.0458(15)$ & $-0.0408(18)$ & 58.55(66) & 54.19(58) & 54.33(59) & 54.29(59)\\
\hline
3 & $0.06618(75)$ & $-0.0523(11)$ & $0.0466(15)$ & $-0.0425(20)$ & 58.28(66) & 53.93(58) & 54.09(58) & 54.04(58)\\
4 & $0.06614(74)$ & $-0.0523(11)$ & $0.0467(15)$ & $-0.0427(19)$ & 58.25(65) & 53.90(57) & 54.06(58) & 54.01(57)\\
5 & $0.06626(74)$ & $-0.0527(11)$ & $0.0473(15)$ & $-0.0438(19)$ & 58.36(65) & 53.99(57) & 54.15(57) & 54.10(57)\\
6 & $0.06829(77)$ & $-0.0557(12)$ & $0.0514(17)$ & $-0.0490(22)$ & 60.14(67) & 55.55(59) & 55.73(59) & 55.67(59)\\
7 & $0.06619(74)$ & $-0.0524(11)$ & $0.0468(15)$ & $-0.0430(19)$ & 58.29(65) & 53.93(57) & 54.10(57) & 54.05(57)\\
8 & $0.06625(74)$ & $-0.0526(11)$ & $0.0470(15)$ & $-0.0429(19)$ & 58.34(65) & 53.98(57) & 54.14(57) & 54.09(57)\\
\hline
9 & $0.06616(77)$ & $-0.0531(12)$ & $0.0483(17)$ & $-0.0450(22)$ & 58.27(68) & 53.87(59) & 54.04(60) & 53.99(59)\\
10 & $0.06630(72)$ & $-0.0534(11)$ & $0.0487(16)$ & $-0.0458(20)$ & 58.39(64) & 53.98(56) & 54.15(56) & 54.10(56)\\
        \end{tabular}
    \end{ruledtabular}
\end{table*}

The power of the Pad\'{e} approximants is illustrated in 
Fig.~\ref{fig:test} which shows the precision of different approximants 
compared with the exact result for a simple test case:
the one-loop quark vacuum polarization function 
from perturbation theory. We set the 
quark mass equal to the kaon mass so that 
the Taylor expansion has the same radius of convergence
as the physical $s$-quark vacuum polarization; this
function also has the same high-$q^2$ behavior as the 
physical function. The Pad\'{e}s converge 
exponentially quickly to the correct result, achieving 
better than 1\% precision after only two terms are included.
It is also clear that the high-$q^2$ contributions are 
accurately approximated by the Pad\'{e}s since
$q>1\,\mathrm{GeV}$, for example, contributes about 1.8\% of
the total $g-2$ correction here. Note also that the
$[2,2]$~approximant is accurate to better than~0.5\% even
when the quark mass is reduced to~$m_\pi$ (as one might do
to simulate $u/d$~vacuum polarization).

The results in Fig.~\ref{fig:test} are for exact moments. 
The finite precision of moments from a simulation obviously
limits the precision of the final results for the anomaly.
The finite precision also limits
the order to which Pad\'{e} approximants can be computed, 
since noisy input data cause spurious poles to appear in
high-order approximants~\cite{Gonnet:2013}\,---\,all poles should be simple, real, 
and located at the poles or on the
branch cut of the exact vacuum polarization function~\cite{Baker:1996}. Higher
orders are possible with more precise moments. The Pad\'{e}s 
typically converge before spurious poles appear, so that the
precision of the final results tracks that of the input moments.
This is 
illustrated in Fig.~\ref{fig:test2} where we have added noise
to the exact moments from one-loop perturbation theory, and compare
the precision of outputs with that of the inputs. Each point
in the plot represents a different simulation, with different noise,
and the colors indicate the order of the approximant used.

As a final check of our analysis method and our simulation codes, 
we generated lattice correlators using our codes but without gauge fields
(that is with link variables $U_\mu(x)=1$), 
and verified (to 0.1\%) that the results for $a_\mu$ agree
with continuum one-loop perturbation theory 
in the limit of zero lattice spacing.

Returning to results from our lattice simulations, 
the Taylor coefficients $\Pi_j$ and contributions to $a_\mu$ from each of
our $s$-quark correlators are shown in Table~\ref{tab:s-data}. 
In each
case results converge to within errors by the $[1,1]$~Pad\'e approximant,
and no spurious poles appear on any of our sets up to and 
including $[2,2]$, 
as expected from our test case. 
Our results on sets 4-7 show that 
finite-volume effects are negligible within 
our 0.1\% statistical errors, but tuning the valence $s$-quark accurately is seen to be important. 

To obtain a final estimate we fit the 
$[2,2]$ results from each configuration set to a function of the form
\begin{align}
      a^s_{\mu,\mathrm{lat}} &= a^s_{\mu} \times \nonumber \\
      &\left( 1 +
      c_{a^2} (a\Lambda_\mathrm{QCD}/\pi)^2
      + c_\mathrm{sea} \delta x_\mathrm{sea} + c_\mathrm{val}\delta x_\mathrm{val}
      \right)
  \end{align}  
where $\Lambda_\mathrm{QCD}=0.5$\,GeV, and $\delta x_\mathrm{sea}$
and $\delta x_\mathrm{val}$ allow for mistuning of the sea and valence
light-quark bare masses:
\begin{align}
    \delta x_\mathrm{sea} &\equiv
    \sum_{q=u,d,s} 
    \frac{m_{q}^\mathrm{sea} - m_{q}^\mathrm{phys}}{m_{s}^\mathrm{phys}}
    \\
    \delta x_s &\equiv
    \frac{m_{s}^\mathrm{val} - m_{s}^\mathrm{phys}}{m_{s}^\mathrm{phys}}.
\end{align}
For our lattices with physical $u/d$ sea masses $\delta x_{\mathrm{sea}}$ is 
very small. $a^2$ errors from staggered `taste-changing' effects will 
remain and they are handled by $c_{a^2}$. 
The four fit parameters are $a_\mu^2$,
$c_{a^2}$, $c_\mathrm{sea}$ and $c_\mathrm{val}$;
we use the following (broad) Gaussian priors for each:
\begin{align}
a_\mu^s &= 0\pm100\times 10^{-10}  \nonumber \\
c_{a^2} &= 0(1) \quad
 c_\mathrm{sea} = 0(1) \quad c_\mathrm{val} = 0(1).
\end{align}

\begin{figure}
   \centering
    \includegraphics[width=0.475\textwidth]{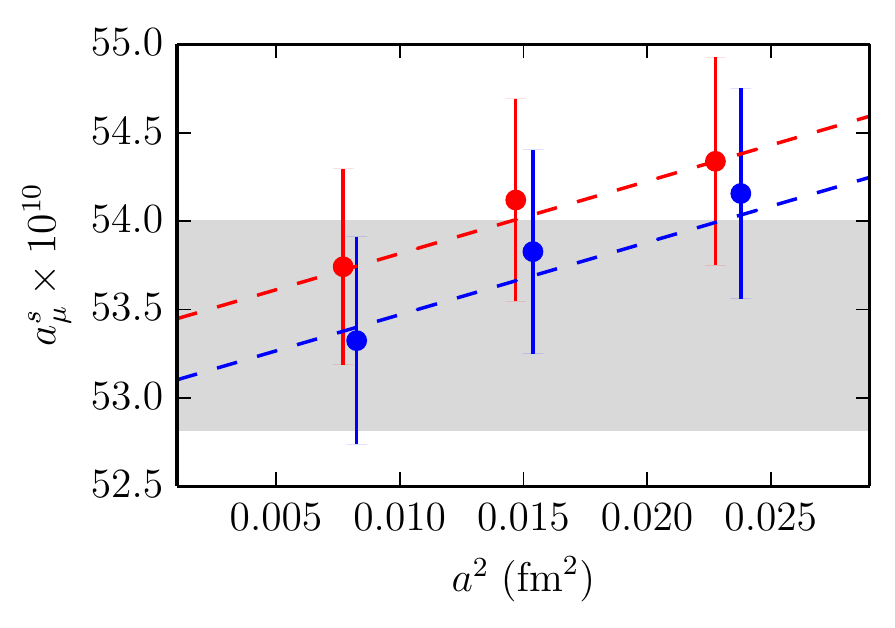}
    \caption{Lattice QCD results for the connected contribution to 
    the muon anomaly $a_\mu$ from vacuum polarization of $s$~quarks. 
    Results are for three lattice
    spacings, and two light-quark masses:
    $m_\ell^\mathrm{lat} = m_s/5$ (lower, blue points), 
    and $m_\ell^\mathrm{lat} = m_\ell^\mathrm{phys}$ (upper, red
    points). The dashed lines are the corresponding values
    from the fit function, with the best-fit parameter values:
    $c_{a^2}=0.29(13)$, $c_\mathrm{sea}=-0.020(6)$ and
    $c_\mathrm{val}=-0.61(4)$. 
    The gray band shows our final result, $53.41(59)\times10^{-10}$, 
    with $m_\ell^\mathrm{lat}=m_\ell^\mathrm{phys}$,
    after extrapolation to $a=0$.
    }
    \label{fig:s-fit}
\end{figure}

\begin{table}
\caption{Error budgets for connected contributions to 
the muon anomaly~$a_\mu$ from vacuum polarization of $s$ and $c$ quarks.}
\label{tab:errbudget}
\begin{ruledtabular}
\begin{tabular}{rcc}
& $a_\mu^s$ & $a_\mu^c$ \\
\hline
Uncertainty in lattice spacing ($w_0$, $r_1$): & 1.0\% & 0.6\% \\
Uncertainty in $Z_V$: & 0.4\% & 2.5\% \\
Monte Carlo statistics: & 0.1\% & 0.1\% \\
$a^2\to0$ extrapolation: & 0.1\% & 0.4\% \\
QED corrections: & 0.1\% & 0.3\% \\
Quark mass tuning: & 0.0\% & 0.4\% \\
Finite lattice volume: & $<0.1\%$ & 0.0\% \\
Pad\'e approximants: & $<0.1\%$ & 0.0\% \\
\hline
Total: & 1.1\% & 2.7\% 
\end{tabular}   
\end{ruledtabular}   
\end{table}

Our final result for the connected contribution for $s$~quarks to $g-2$ is:
\begin{equation}
    a_\mu^s = 53.41(59) \times 10^{-10}.
\end{equation}
The fit to $[2,2]$ Pad\'e results from all~10 of our configuration sets is
excellent, with a $\chi^2$ per degree of freedom of 0.22 ($p$-value of 0.99). 
In Fig.~\ref{fig:s-fit} we compare our fit with the data from configurations with
$m_s/m_\ell$ equal~5 and with the physical mass ratio. 

\begin{table}
\caption{Contributions to $a_\mu$ from $s$~and $c$~quark vacuum polarization.
Only connected parts of the vacuum polarization are included.
Results, multiplied by $10^{10}$, 
are shown for each of the Pad\'e approximants.
}
\label{tab:sc-results}
\begin{ruledtabular}
\begin{tabular}{ccccc}
    Quark 
    & $[1,0]\times10^{10}$ & $[1,1]\times10^{10}$ 
    & $[2,1]\times10^{10}$ & $[2,2]\times10^{10}$
    \\  \hline
    $s$ & 57.63(67) & 53.28(58) & 53.46(59) & 53.41(59) \\
    $c$ & 14.58(39) & 14.41(39) & 14.42(39) & 14.42(39)
\end{tabular}
\end{ruledtabular}
\end{table}

The error budget for 
our result is given in Table~\ref{tab:errbudget}. The dominant error, by
far, comes from the uncertainty in the physical value 
of the Wilson flow parameter~$w_0$, 
which we use to set the lattice spacings. We estimate the uncertainty from
QED corrections to the vacuum polarization to be of order~0.1\% from 
perturbation theory~\cite{Donald:2012ga}, 
suppressed by the small charge of the $s$~quark. Our results show negligible
dependence ($<0.1$\%) on the spatial size of the lattice, which we varied by a factor 
of two. Also the convergence of successive orders of Pad\'e approximant indicates
convergence to better than~0.1\%; results from fits to different approximants
are tabulated in Table~\ref{tab:sc-results}. 

Note that the $a^2$ errors are quite small in our analysis. 
This is because we use the highly
corrected HISQ discretization of the quark action. Our final ($a=0$) result
is only 0.6\% below our results from the 0.09\,fm lattices (sets~9 and~10). 
The variation from our coarsest lattice to $a=0$ is only 1.8\%. We compared
this with results from the clover discretization for quarks, 
which had finite-$a$ errors in excess of 20\% on the coarsest lattices.

Finally we also include results for $c$~quarks in Tables~\ref{tab:errbudget}
and~\ref{tab:sc-results}. These are calculated from the moments (and
error budget) published in~\cite{Donald:2012ga}.  Our final result for the
connected contribution to the muon anomaly from $c$-quark vacuum 
polarization is:
\begin{equation}
    a_\mu^c = 14.42(39)\times 10^{-10}.
\end{equation}
The dominant source of error here is in the determination of the $Z_V$~renormalization
factors. This error could be substantially reduced by using the method
we used for the $s$-quark contribution~\cite{Chakraborty:2014zma}.

\section{Discussion/Conclusions}
\label{sec:discussion}

The ultimate aim of lattice QCD calculations of $a_{\mu,\mathrm{HVP}}$ is 
to improve on results from using, for example, $\sigma(e^+e^- \rightarrow \mathrm{hadrons})$ 
that are able to achieve an uncertainty of below 1\%. 
We are not at that stage yet. However, our results here show 
that a 1\% error can be achieved now for the connected piece of the $s$-quark 
contribution and a 1\% error could easily be achieved for the 
$c$-quark. It then makes sense to try to separate out the 
$s$-quark or $c$-quark piece of the result from $\sigma_{e^+e^-}$ for comparison. 
The flavour identification is not completely unambiguous in that case nor
can disconnected contributions or QED effects be removed, so the accuracy 
with which this can be done is limited. 

For the $c$-quark contribution a result of $14.4(1)\times 10^{-10}$ is given 
by~\cite{Bodenstein}, with which our result is in good agreement. 
For the $s$-quark contribution, we can make predictions based on the  
data compilation from~\cite{Hagiwara:2011af} for energies up to $2$ GeV in 
the dispersion integral and by using perturbative QCD for higher energies. 
The leading contribution can be estimated from the sum of the 
$K^+K^-$ and $K_L^0K_S^0$ channels, adding up to $35.5\times 10^{-10}$. 
As most of this contribution is from the $\phi$ resonance, we can assume 
that this is included in the lattice calculation. It is less clear how 
much of the remaining $\phi$ decay channels should be included for the 
comparison, as these may require the inclusion of disconnected diagrams 
in the lattice calculation. 
However, using the branching 
fractions for $\phi\to K^+K^-$ and $\phi\to K_L^0K_S^0$ 
we can predict a maximum $\phi$ contribution of $42.8\times 10^{-10}$. 
Other channels containing $\overline{K}K$ ($\overline{K}Kn\pi$, $\overline{K}K\omega$) 
and $\eta\phi$, which are taken into account in~\cite{Hagiwara:2011af}, add up 
to $6.6\times 10^{-10}$. It is not clear to what extent these are included 
in the lattice calculation presented here and to what extent they derive 
from a light quark loop coupled to the electromagnetic current together 
with $s$ sea quarks. The $s$ quark contribution for energies above 
2 GeV can be reliably calculated from 
perturbative QCD. It is $5.9 \times 10^{-10}$ and must certainly be included 
in the comparison. With these numbers we predict the total 
$s$ quark contribution from data and perturbative QCD to be 
$55.3(8)\times 10^{-10}$. From the caveats mentioned, this should 
be seen as an upper limit for the comparison with lattice QCD.
Previous lattice QCD calculations have not reported separate results for 
$s$ or $c$ quarks (if calculated), despite the simplicity of doing this when 
only the connected contribution has been calculated. It would seem sensible to do this
for comparison of different lattice results in the future. 

The extension of our method for light-quark ($u,d$) vacuum polarization is straightforward, with one modification. The light-quark contribution is the most important in the total HVP, being about 12 times larger than that for the strange quark, in part because of a factor of~5 from the electric charges.
The one complication is that the signal-to-noise 
at large $t$ is much worse for the light-quark correlators (because $2m_\pi$ is small compared to $m_\rho$), 
greatly increasing the statistical errors of the moments. 
This is easily handled by fitting Monte Carlo data for the correlator $G(t)$ with a standard multi-exponential fit function $G_\mathrm{fit}(t,p)$, 
where the $p_\beta$ are fit parameters. 
The moments are then calculated from $G_\mathrm{fit}(t,p^*)$, using the best-fit parameter values~$p=p^*$, rather than from the data. The fit function has similar
errors to the data at low~$t$, but much smaller errors (orders of magnitude) 
at large~$t$, and therefore much smaller errors in the moments. By 
using the fit function, we build into our $g-2$ analysis
knowledge about how  $G(t)$'s behavior at large~$t$ is constrained
by its behavior at small~$t$.

We tested this fitting idea on our $s$-quark data, using data for $t=0,\pm1$ 
and the best-fit function ($G_\mathrm{fit}$) for all other $t$ when computing
moments. We obtained results identical, to four decimal places or better, 
with what we found above (Table~\ref{tab:s-data}) and with slightly smaller statistical 
errors. 
We also tested this idea on 
a single low-statistics sample of correlators 
from 4 time sources on 400 configurations (a subset of 
Set~8 of Table~\ref{tab:params}) 
with the valence quark mass equal to 
the physical light-quark mass $m_\ell^\mathrm{phys}$. 
Using $G_\mathrm{fit}(t,p^*)$
in place of the $G(t)$ reduced the statistical errors 
from $\pm100\%$ to~$\pm6\%$, indicating that errors of around 3\% might 
be achieved using the full statistics. 
As expected the Pad\'e approximants converged to better than~1\% by
the [1,1]~approximant.
We will discuss light quarks in a separate paper, but an uncertainty of 1\%, 
as achieved here for $s$ quarks, seems feasible on ensemble sizes
of ~$10\times$ that used here. 

Using $G_\mathrm{fit}(t)$ to calculate moments also allows us to remove
systematic errors caused by the finite temporal size~$T$ of our 
lattices. This is because it is trivial to take $T\to\infty$
in $G_\mathrm{fit}(t)$ after fitting but before
calculating the moments. The resulting shifts in $a_\mu$ are 
typically very small\,---for example, \,only~0.09\% for
the physical $u/d$~quark propagators in our test analysis above\,---\,but
the correction is worth making anyway because it is so simple.

Our method also provides a straightforward extension to include disconnected 
contributions. All that is necessary is to calculate the disconnected contribution to 
the vector correlator at zero spatial momentum and that can be done with existing 
techniques, again provided adequate statistics are available   
\footnote{Note that, while 
in the $SU(3)$ limit the disconnected contributions would vanish, negative 
contributions of about 9\% have been predicted based on chiral perturbation 
theory~\cite{DellaMorte}.}. 
To reduce errors below 1\% on the total HVP contribution 
may require direct calculation of QED effects on the 
lattice and the incorporation of $u$ and $d$ quark propagators of different mass, techniques that 
are being tested now in the lattice QCD community.  

In conclusion, we have shown that a simple method using a small number of time-moments of 
vector current-current correlators can yield 1\% accurate results for the hadronic 
vacuum polarization contribution to $a_{\mu}$. For the connected $s$-quark 
and $c$-quark contributions we find, respectively, $a_{\mu}^s = 53.41(59)\times 10^{-10}$ 
and $a_{\mu}^c = 14.42(39) \times 10^{-10}$. 

{\it Note added.} After our paper appeared we received the results 
of a preliminary analysis by members of the ETM Collaboration~\cite{priv}, 
separating $a_{\mu}^s$ and $a_{\mu}^c$ from their analysis of the complete 
4-flavour connected HVP contribution to $a_{\mu}$~\cite{Burger:2013jya}. 
They find $a_{\mu}^s = 53(3) \times 10^{-10}$ 
and $a_{\mu}^c = 14.1(6) \times 10^{-10}$, in agreement with our 
results. 

{\it Acknowledgements.} We are grateful to the MILC collaboration 
for the use of their 
gauge configurations and code. 
We thank R.~R.~Horgan, G.~M.~von~Hippel, D.~Nomura, P.~Rakow and D.~Toussaint for useful conversations.
Our calculations were done on the Darwin Supercomputer 
as part of STFC's DiRAC facility jointly
funded by STFC, BIS 
and the Universities of Cambridge and Glasgow. 
This work was funded by STFC, the Royal Society, the Wolfson Foundation
and the National Science Foundation.
\bibliography{g2s}

\begin{thebibliography}{39}
\expandafter\ifx\csname natexlab\endcsname\relax\def\natexlab#1{#1}\fi
\expandafter\ifx\csname bibnamefont\endcsname\relax
  \def\bibnamefont#1{#1}\fi
\expandafter\ifx\csname bibfnamefont\endcsname\relax
  \def\bibfnamefont#1{#1}\fi
\expandafter\ifx\csname citenamefont\endcsname\relax
  \def\citenamefont#1{#1}\fi
\expandafter\ifx\csname url\endcsname\relax
  \def\url#1{\texttt{#1}}\fi
\expandafter\ifx\csname urlprefix\endcsname\relax\def\urlprefix{URL }\fi
\providecommand{\bibinfo}[2]{#2}
\providecommand{\eprint}[2][]{\url{#2}}

\bibitem[{\citenamefont{Bennett et~al.}(2006)}]{Bennett:2006fi}
\bibinfo{author}{\bibfnamefont{G.}~\bibnamefont{Bennett}} \bibnamefont{et~al.}
  (\bibinfo{collaboration}{Muon G-2 Collaboration}),
  \bibinfo{journal}{Phys.Rev.} \textbf{\bibinfo{volume}{D73}},
  \bibinfo{pages}{072003} (\bibinfo{year}{2006}), \eprint{hep-ex/0602035}.

\bibitem[{\citenamefont{Aoyama et~al.}(2012)\citenamefont{Aoyama, Hayakawa,
  Kinoshita, and Nio}}]{Aoyama:2012wk}
\bibinfo{author}{\bibfnamefont{T.}~\bibnamefont{Aoyama}},
  \bibinfo{author}{\bibfnamefont{M.}~\bibnamefont{Hayakawa}},
  \bibinfo{author}{\bibfnamefont{T.}~\bibnamefont{Kinoshita}},
  \bibnamefont{and} \bibinfo{author}{\bibfnamefont{M.}~\bibnamefont{Nio}},
  \bibinfo{journal}{Phys.Rev.Lett.} \textbf{\bibinfo{volume}{109}},
  \bibinfo{pages}{111808} (\bibinfo{year}{2012}), \eprint{1205.5370}.

\bibitem[{\citenamefont{Hagiwara et~al.}(2011)\citenamefont{Hagiwara, Liao,
  Martin, Nomura, and Teubner}}]{Hagiwara:2011af}
\bibinfo{author}{\bibfnamefont{K.}~\bibnamefont{Hagiwara}},
  \bibinfo{author}{\bibfnamefont{R.}~\bibnamefont{Liao}},
  \bibinfo{author}{\bibfnamefont{A.~D.} \bibnamefont{Martin}},
  \bibinfo{author}{\bibfnamefont{D.}~\bibnamefont{Nomura}}, \bibnamefont{and}
  \bibinfo{author}{\bibfnamefont{T.}~\bibnamefont{Teubner}},
  \bibinfo{journal}{J.Phys.} \textbf{\bibinfo{volume}{G38}},
  \bibinfo{pages}{085003} (\bibinfo{year}{2011}), \eprint{1105.3149}.

\bibitem[{\citenamefont{Davier et~al.}(2011)\citenamefont{Davier, Hoecker,
  Malaescu, and Zhang}}]{Davier}
\bibinfo{author}{\bibfnamefont{M.}~\bibnamefont{Davier}},
  \bibinfo{author}{\bibfnamefont{A.}~\bibnamefont{Hoecker}},
  \bibinfo{author}{\bibfnamefont{B.}~\bibnamefont{Malaescu}}, \bibnamefont{and}
  \bibinfo{author}{\bibfnamefont{Z.}~\bibnamefont{Zhang}},
  \bibinfo{journal}{Eur.Phys.J.} \textbf{\bibinfo{volume}{C71}},
  \bibinfo{pages}{1515} (\bibinfo{year}{2011}), \eprint{1010.4180}.

\bibitem[{\citenamefont{Blum et~al.}(2013)\citenamefont{Blum, Denig,
  Logashenko, de~Rafael, Lee~Roberts et~al.}}]{snowmass}
\bibinfo{author}{\bibfnamefont{T.}~\bibnamefont{Blum}},
  \bibinfo{author}{\bibfnamefont{A.}~\bibnamefont{Denig}},
  \bibinfo{author}{\bibfnamefont{I.}~\bibnamefont{Logashenko}},
  \bibinfo{author}{\bibfnamefont{E.}~\bibnamefont{de~Rafael}},
  \bibinfo{author}{\bibfnamefont{B.}~\bibnamefont{Lee~Roberts}},
  \bibnamefont{et~al.} (\bibinfo{year}{2013}), \eprint{1311.2198}.

\bibitem[{\citenamefont{Blum}(2003)}]{Blum:2002ii}
\bibinfo{author}{\bibfnamefont{T.}~\bibnamefont{Blum}},
  \bibinfo{journal}{Phys.Rev.Lett.} \textbf{\bibinfo{volume}{91}},
  \bibinfo{pages}{052001} (\bibinfo{year}{2003}), \eprint{hep-lat/0212018}.

\bibitem[{\citenamefont{Blum et~al.}(2012)\citenamefont{Blum, Hayakawa, and
  Izubuchi}}]{Blum:2013qu}
\bibinfo{author}{\bibfnamefont{T.}~\bibnamefont{Blum}},
  \bibinfo{author}{\bibfnamefont{M.}~\bibnamefont{Hayakawa}}, \bibnamefont{and}
  \bibinfo{author}{\bibfnamefont{T.}~\bibnamefont{Izubuchi}},
  \bibinfo{journal}{PoS} \textbf{\bibinfo{volume}{LATTICE2012}},
  \bibinfo{pages}{022} (\bibinfo{year}{2012}), \eprint{1301.2607}.

\bibitem[{\citenamefont{Aubin and Blum}(2007)}]{aubin-blum}
\bibinfo{author}{\bibfnamefont{C.}~\bibnamefont{Aubin}} \bibnamefont{and}
  \bibinfo{author}{\bibfnamefont{T.}~\bibnamefont{Blum}},
  \bibinfo{journal}{Phys.Rev.} \textbf{\bibinfo{volume}{D75}},
  \bibinfo{pages}{114502} (\bibinfo{year}{2007}), \eprint{hep-lat/0608011}.

\bibitem[{\citenamefont{Boyle et~al.}(2012)\citenamefont{Boyle, Del~Debbio,
  Kerrane, and Zanotti}}]{boyle}
\bibinfo{author}{\bibfnamefont{P.}~\bibnamefont{Boyle}},
  \bibinfo{author}{\bibfnamefont{L.}~\bibnamefont{Del~Debbio}},
  \bibinfo{author}{\bibfnamefont{E.}~\bibnamefont{Kerrane}}, \bibnamefont{and}
  \bibinfo{author}{\bibfnamefont{J.}~\bibnamefont{Zanotti}},
  \bibinfo{journal}{Phys.Rev.} \textbf{\bibinfo{volume}{D85}},
  \bibinfo{pages}{074504} (\bibinfo{year}{2012}), \eprint{1107.1497}.

\bibitem[{\citenamefont{Feng et~al.}(2011)\citenamefont{Feng, Jansen,
  Petschlies, and Renner}}]{renner}
\bibinfo{author}{\bibfnamefont{X.}~\bibnamefont{Feng}},
  \bibinfo{author}{\bibfnamefont{K.}~\bibnamefont{Jansen}},
  \bibinfo{author}{\bibfnamefont{M.}~\bibnamefont{Petschlies}},
  \bibnamefont{and} \bibinfo{author}{\bibfnamefont{D.~B.}
  \bibnamefont{Renner}}, \bibinfo{journal}{Phys.Rev.Lett.}
  \textbf{\bibinfo{volume}{107}}, \bibinfo{pages}{081802}
  (\bibinfo{year}{2011}), \eprint{1103.4818}.

\bibitem[{\citenamefont{Della~Morte et~al.}(2012)\citenamefont{Della~Morte,
  Jager, Juttner, and Wittig}}]{wittig}
\bibinfo{author}{\bibfnamefont{M.}~\bibnamefont{Della~Morte}},
  \bibinfo{author}{\bibfnamefont{B.}~\bibnamefont{Jager}},
  \bibinfo{author}{\bibfnamefont{A.}~\bibnamefont{Juttner}}, \bibnamefont{and}
  \bibinfo{author}{\bibfnamefont{H.}~\bibnamefont{Wittig}},
  \bibinfo{journal}{JHEP} \textbf{\bibinfo{volume}{1203}}, \bibinfo{pages}{055}
  (\bibinfo{year}{2012}), \eprint{1112.2894}.

\bibitem[{\citenamefont{Aubin et~al.}(2013)\citenamefont{Aubin, Blum,
  Golterman, and Peris}}]{aubin-twisted}
\bibinfo{author}{\bibfnamefont{C.}~\bibnamefont{Aubin}},
  \bibinfo{author}{\bibfnamefont{T.}~\bibnamefont{Blum}},
  \bibinfo{author}{\bibfnamefont{M.}~\bibnamefont{Golterman}},
  \bibnamefont{and} \bibinfo{author}{\bibfnamefont{S.}~\bibnamefont{Peris}},
  \bibinfo{journal}{Phys.Rev.} \textbf{\bibinfo{volume}{D88}},
  \bibinfo{pages}{074505} (\bibinfo{year}{2013}), \eprint{1307.4701}.

\bibitem[{\citenamefont{Aubin et~al.}(2012)\citenamefont{Aubin, Blum,
  Golterman, and Peris}}]{Aubin:2012me}
\bibinfo{author}{\bibfnamefont{C.}~\bibnamefont{Aubin}},
  \bibinfo{author}{\bibfnamefont{T.}~\bibnamefont{Blum}},
  \bibinfo{author}{\bibfnamefont{M.}~\bibnamefont{Golterman}},
  \bibnamefont{and} \bibinfo{author}{\bibfnamefont{S.}~\bibnamefont{Peris}},
  \bibinfo{journal}{Phys.Rev.} \textbf{\bibinfo{volume}{D86}},
  \bibinfo{pages}{054509} (\bibinfo{year}{2012}), \eprint{1205.3695}.

\bibitem[{\citenamefont{Feng et~al.}(2013)\citenamefont{Feng, Hashimoto,
  Hotzel, Jansen, Petschlies et~al.}}]{Feng:2013xsa}
\bibinfo{author}{\bibfnamefont{X.}~\bibnamefont{Feng}},
  \bibinfo{author}{\bibfnamefont{S.}~\bibnamefont{Hashimoto}},
  \bibinfo{author}{\bibfnamefont{G.}~\bibnamefont{Hotzel}},
  \bibinfo{author}{\bibfnamefont{K.}~\bibnamefont{Jansen}},
  \bibinfo{author}{\bibfnamefont{M.}~\bibnamefont{Petschlies}},
  \bibnamefont{et~al.}, \bibinfo{journal}{Phys.Rev.}
  \textbf{\bibinfo{volume}{D88}}, \bibinfo{pages}{034505}
  (\bibinfo{year}{2013}), \eprint{1305.5878}.

\bibitem[{\citenamefont{Francis et~al.}(2013)\citenamefont{Francis, Jaeger,
  Meyer, and Wittig}}]{Francis:2013fzp}
\bibinfo{author}{\bibfnamefont{A.}~\bibnamefont{Francis}},
  \bibinfo{author}{\bibfnamefont{B.}~\bibnamefont{Jaeger}},
  \bibinfo{author}{\bibfnamefont{H.~B.} \bibnamefont{Meyer}}, \bibnamefont{and}
  \bibinfo{author}{\bibfnamefont{H.}~\bibnamefont{Wittig}},
  \bibinfo{journal}{Phys.Rev.} \textbf{\bibinfo{volume}{D88}},
  \bibinfo{pages}{054502} (\bibinfo{year}{2013}), \eprint{1306.2532}.

\bibitem[{\citenamefont{Gregory et~al.}(2013)\citenamefont{Gregory, Fodor,
  Hoelbling, Krieg, Lellouch et~al.}}]{Gregory:2013taa}
\bibinfo{author}{\bibfnamefont{E.~B.} \bibnamefont{Gregory}},
  \bibinfo{author}{\bibfnamefont{Z.}~\bibnamefont{Fodor}},
  \bibinfo{author}{\bibfnamefont{C.}~\bibnamefont{Hoelbling}},
  \bibinfo{author}{\bibfnamefont{S.}~\bibnamefont{Krieg}},
  \bibinfo{author}{\bibfnamefont{L.}~\bibnamefont{Lellouch}},
  \bibnamefont{et~al.} (\bibinfo{year}{2013}), \eprint{1311.4446}.

\bibitem[{Pad()}]{Pade}
\bibinfo{note}{{The $[m,n]$ Pad\'e approximant of a function $f(x)$ is a ratio
  of polynomials in~$x$, of order $m$ in the numerator and $n$ in the
  denominator, whose Taylor expansion is the same as that of $f(x)$ through
  order~$x^{n+m}$.}}

\bibitem[{\citenamefont{Baker}(1969)}]{Baker:1969}
\bibinfo{author}{\bibfnamefont{G.}~\bibnamefont{Baker}},
  \bibinfo{journal}{J.Math.Phys.} \textbf{\bibinfo{volume}{10}},
  \bibinfo{pages}{814} (\bibinfo{year}{1969}),
  \eprint{http://dx.doi.org/10.1063/1.1664911}.

\bibitem[{\citenamefont{Baker and Graves-Morris}(1996)}]{Baker:1996}
\bibinfo{author}{\bibfnamefont{G.}~\bibnamefont{Baker}} \bibnamefont{and}
  \bibinfo{author}{\bibfnamefont{P.}~\bibnamefont{Graves-Morris}},
  \emph{\bibinfo{title}{Pad\'e Approximants}} (\bibinfo{publisher}{Cambridge},
  \bibinfo{year}{1996}), \bibinfo{edition}{2nd} ed., \bibinfo{note}{especially
  chapter~5}.

\bibitem[{\citenamefont{Donald et~al.}(2012)\citenamefont{Donald, Davies,
  Dowdall, Follana, Hornbostel et~al.}}]{Donald:2012ga}
\bibinfo{author}{\bibfnamefont{G.}~\bibnamefont{Donald}},
  \bibinfo{author}{\bibfnamefont{C.}~\bibnamefont{Davies}},
  \bibinfo{author}{\bibfnamefont{R.}~\bibnamefont{Dowdall}},
  \bibinfo{author}{\bibfnamefont{E.}~\bibnamefont{Follana}},
  \bibinfo{author}{\bibfnamefont{K.}~\bibnamefont{Hornbostel}},
  \bibnamefont{et~al.} (\bibinfo{collaboration}{HPQCD Collaboration}),
  \bibinfo{journal}{Phys.Rev.} \textbf{\bibinfo{volume}{D86}},
  \bibinfo{pages}{094501} (\bibinfo{year}{2012}), \eprint{1208.2855}.

\bibitem[{\citenamefont{Davies et~al.}(2012)\citenamefont{Davies, Donald,
  Dowdall, Koponen, Follana et~al.}}]{jpsi}
\bibinfo{author}{\bibfnamefont{C.}~\bibnamefont{Davies}},
  \bibinfo{author}{\bibfnamefont{G.}~\bibnamefont{Donald}},
  \bibinfo{author}{\bibfnamefont{R.}~\bibnamefont{Dowdall}},
  \bibinfo{author}{\bibfnamefont{J.}~\bibnamefont{Koponen}},
  \bibinfo{author}{\bibfnamefont{E.}~\bibnamefont{Follana}},
  \bibnamefont{et~al.} (\bibinfo{collaboration}{HPQCD Collaboration}),
  \bibinfo{journal}{PoS} \textbf{\bibinfo{volume}{ConfinementX}},
  \bibinfo{pages}{288} (\bibinfo{year}{2012}), \eprint{1301.7203}.

\bibitem[{\citenamefont{Bazavov et~al.}(2010)}]{Bazavov:2010ru}
\bibinfo{author}{\bibfnamefont{A.}~\bibnamefont{Bazavov}} \bibnamefont{et~al.}
  (\bibinfo{collaboration}{MILC collaboration}), \bibinfo{journal}{Phys.Rev.}
  \textbf{\bibinfo{volume}{D82}}, \bibinfo{pages}{074501}
  (\bibinfo{year}{2010}), \eprint{1004.0342}.

\bibitem[{\citenamefont{Bazavov et~al.}(2013)}]{Bazavov:2012uw}
\bibinfo{author}{\bibfnamefont{A.}~\bibnamefont{Bazavov}} \bibnamefont{et~al.}
  (\bibinfo{collaboration}{MILC Collaboration}), \bibinfo{journal}{Phys.Rev.}
  \textbf{\bibinfo{volume}{D87}}, \bibinfo{pages}{054505}
  (\bibinfo{year}{2013}), \eprint{1212.4768}.

\bibitem[{\citenamefont{Dowdall et~al.}(2013)\citenamefont{Dowdall, Davies,
  Lepage, and McNeile}}]{fkpi}
\bibinfo{author}{\bibfnamefont{R.}~\bibnamefont{Dowdall}},
  \bibinfo{author}{\bibfnamefont{C.}~\bibnamefont{Davies}},
  \bibinfo{author}{\bibfnamefont{G.}~\bibnamefont{Lepage}}, \bibnamefont{and}
  \bibinfo{author}{\bibfnamefont{C.}~\bibnamefont{McNeile}}
  (\bibinfo{collaboration}{HPQCD Collaboration}), \bibinfo{journal}{Phys.Rev.}
  \textbf{\bibinfo{volume}{D88}}, \bibinfo{pages}{074504}
  (\bibinfo{year}{2013}), \eprint{1303.1670}.

\bibitem[{\citenamefont{Borsanyi et~al.}(2012)\citenamefont{Borsanyi, Durr,
  Fodor, Hoelbling, Katz et~al.}}]{Borsanyi:2012zs}
\bibinfo{author}{\bibfnamefont{S.}~\bibnamefont{Borsanyi}},
  \bibinfo{author}{\bibfnamefont{S.}~\bibnamefont{Durr}},
  \bibinfo{author}{\bibfnamefont{Z.}~\bibnamefont{Fodor}},
  \bibinfo{author}{\bibfnamefont{C.}~\bibnamefont{Hoelbling}},
  \bibinfo{author}{\bibfnamefont{S.~D.} \bibnamefont{Katz}},
  \bibnamefont{et~al.}, \bibinfo{journal}{JHEP}
  \textbf{\bibinfo{volume}{1209}}, \bibinfo{pages}{010} (\bibinfo{year}{2012}),
  \eprint{1203.4469}.

\bibitem[{\citenamefont{Chakraborty et~al.}(2013)\citenamefont{Chakraborty,
  Davies, Donald, Dowdall, Koponen et~al.}}]{Chakraborty:2014zma}
\bibinfo{author}{\bibfnamefont{B.}~\bibnamefont{Chakraborty}},
  \bibinfo{author}{\bibfnamefont{C.}~\bibnamefont{Davies}},
  \bibinfo{author}{\bibfnamefont{G.}~\bibnamefont{Donald}},
  \bibinfo{author}{\bibfnamefont{R.}~\bibnamefont{Dowdall}},
  \bibinfo{author}{\bibfnamefont{J.}~\bibnamefont{Koponen}},
  \bibnamefont{et~al.} (\bibinfo{collaboration}{HPQCD Collaboration}),
  \bibinfo{journal}{PoS} \textbf{\bibinfo{volume}{LATTICE2013}},
  \bibinfo{pages}{309} (\bibinfo{year}{2013}), \eprint{1401.0669}.

\bibitem[{\citenamefont{Follana et~al.}(2007)\citenamefont{Follana, Mason,
  Davies, Hornbostel, Lepage et~al.}}]{HISQ_PRD}
\bibinfo{author}{\bibfnamefont{E.}~\bibnamefont{Follana}},
  \bibinfo{author}{\bibfnamefont{Q.}~\bibnamefont{Mason}},
  \bibinfo{author}{\bibfnamefont{C.}~\bibnamefont{Davies}},
  \bibinfo{author}{\bibfnamefont{K.}~\bibnamefont{Hornbostel}},
  \bibinfo{author}{\bibfnamefont{G.~P.} \bibnamefont{Lepage}},
  \bibnamefont{et~al.} (\bibinfo{collaboration}{HPQCD and UKQCD
  Collaborations}), \bibinfo{journal}{Phys.Rev.}
  \textbf{\bibinfo{volume}{D75}}, \bibinfo{pages}{054502}
  (\bibinfo{year}{2007}), \eprint{hep-lat/0610092}.

\bibitem[{\citenamefont{Follana et~al.}(2008)\citenamefont{Follana, Davies,
  Lepage, and Shigemitsu}}]{HISQ_PRL}
\bibinfo{author}{\bibfnamefont{E.}~\bibnamefont{Follana}},
  \bibinfo{author}{\bibfnamefont{C.}~\bibnamefont{Davies}},
  \bibinfo{author}{\bibfnamefont{G.}~\bibnamefont{Lepage}}, \bibnamefont{and}
  \bibinfo{author}{\bibfnamefont{J.}~\bibnamefont{Shigemitsu}}
  (\bibinfo{collaboration}{HPQCD and UKQCD Collaborations}),
  \bibinfo{journal}{Phys.Rev.Lett.} \textbf{\bibinfo{volume}{100}},
  \bibinfo{pages}{062002} (\bibinfo{year}{2008}), \eprint{0706.1726}.

\bibitem[{\citenamefont{Davies et~al.}(2010{\natexlab{a}})\citenamefont{Davies,
  McNeile, Follana, Lepage, Na et~al.}}]{Dsdecayconst}
\bibinfo{author}{\bibfnamefont{C.}~\bibnamefont{Davies}},
  \bibinfo{author}{\bibfnamefont{C.}~\bibnamefont{McNeile}},
  \bibinfo{author}{\bibfnamefont{E.}~\bibnamefont{Follana}},
  \bibinfo{author}{\bibfnamefont{G.}~\bibnamefont{Lepage}},
  \bibinfo{author}{\bibfnamefont{H.}~\bibnamefont{Na}}, \bibnamefont{et~al.}
  (\bibinfo{collaboration}{HPQCD Collaboration}), \bibinfo{journal}{Phys.Rev.}
  \textbf{\bibinfo{volume}{D82}}, \bibinfo{pages}{114504}
  (\bibinfo{year}{2010}{\natexlab{a}}), \eprint{1008.4018}.

\bibitem[{\citenamefont{Durr et~al.}(2011)\citenamefont{Durr, Fodor, Hoelbling,
  Katz, Krieg et~al.}}]{durrmsml}
\bibinfo{author}{\bibfnamefont{S.}~\bibnamefont{Durr}},
  \bibinfo{author}{\bibfnamefont{Z.}~\bibnamefont{Fodor}},
  \bibinfo{author}{\bibfnamefont{C.}~\bibnamefont{Hoelbling}},
  \bibinfo{author}{\bibfnamefont{S.}~\bibnamefont{Katz}},
  \bibinfo{author}{\bibfnamefont{S.}~\bibnamefont{Krieg}},
  \bibnamefont{et~al.}, \bibinfo{journal}{Phys.Lett.}
  \textbf{\bibinfo{volume}{B701}}, \bibinfo{pages}{265} (\bibinfo{year}{2011}),
  \eprint{1011.2403}.

\bibitem[{\citenamefont{Davies et~al.}(2010{\natexlab{b}})\citenamefont{Davies,
  Follana, Kendall, Lepage, and McNeile}}]{Davies:2009tsa}
\bibinfo{author}{\bibfnamefont{C.}~\bibnamefont{Davies}},
  \bibinfo{author}{\bibfnamefont{E.}~\bibnamefont{Follana}},
  \bibinfo{author}{\bibfnamefont{I.}~\bibnamefont{Kendall}},
  \bibinfo{author}{\bibfnamefont{G.}~\bibnamefont{Lepage}}, \bibnamefont{and}
  \bibinfo{author}{\bibfnamefont{C.}~\bibnamefont{McNeile}}
  (\bibinfo{collaboration}{HPQCD Collaboration}), \bibinfo{journal}{Phys.Rev.}
  \textbf{\bibinfo{volume}{D81}}, \bibinfo{pages}{034506}
  (\bibinfo{year}{2010}{\natexlab{b}}), \eprint{0910.1229}.

\bibitem[{dis()}]{disc}
\bibinfo{note}{{Eq.~(\ref{eq:G}) is simply related to finite-difference
  approximations of derivatives of $\hat\Pi(q^2)$ that use the discrete values
  of $q=E_n$ allowed by the periodic $t$-lattice. The simplest
  finite-difference derivative, $\Delta f(E_n) \equiv
  (f(E_{n+1})-f(E_{n-1}))/2\delta E$ where $\delta E=E_{n+1}-E_n$, is accurate
  up to an error of order $(\delta E)^2$. This and higher-order errors can be
  removed by making successively less local approximations: \[\Delta - (\delta
  E)^2 \Delta^3/6 + 3(\delta E)^4 \Delta^5/40 + \cdots.\] Continuing this
  process to remove errors to all orders in $(\delta E)^2$ results in the
  derivative defined by Eq.~(\ref{eq:G}), as is evident from the Fourier
  transform of the improved difference operator.}}

\bibitem[{\citenamefont{Allison et~al.}(2008)}]{Allison:2008xk}
\bibinfo{author}{\bibfnamefont{I.}~\bibnamefont{Allison}} \bibnamefont{et~al.}
  (\bibinfo{collaboration}{HPQCD Collaboration}), \bibinfo{journal}{Phys.Rev.}
  \textbf{\bibinfo{volume}{D78}}, \bibinfo{pages}{054513}
  (\bibinfo{year}{2008}), \eprint{0805.2999}.

\bibitem[{\citenamefont{McNeile et~al.}(2010)\citenamefont{McNeile, Davies,
  Follana, Hornbostel, and Lepage}}]{McNeile:2010ji}
\bibinfo{author}{\bibfnamefont{C.}~\bibnamefont{McNeile}},
  \bibinfo{author}{\bibfnamefont{C.}~\bibnamefont{Davies}},
  \bibinfo{author}{\bibfnamefont{E.}~\bibnamefont{Follana}},
  \bibinfo{author}{\bibfnamefont{K.}~\bibnamefont{Hornbostel}},
  \bibnamefont{and} \bibinfo{author}{\bibfnamefont{G.}~\bibnamefont{Lepage}}
  (\bibinfo{collaboration}{HPQCD Collaboration}), \bibinfo{journal}{Phys.Rev.}
  \textbf{\bibinfo{volume}{D82}}, \bibinfo{pages}{034512}
  (\bibinfo{year}{2010}), \eprint{1004.4285}.

\bibitem[{\citenamefont{Gonnet et~al.}(2013)}]{Gonnet:2013}
\bibinfo{author}{\bibfnamefont{P.}~\bibnamefont{Gonnet}} \bibnamefont{et~al.},
  \bibinfo{journal}{SIAM Review} \textbf{\bibinfo{volume}{55}},
  \bibinfo{pages}{101} (\bibinfo{year}{2013}).

\bibitem[{\citenamefont{Bodenstein et~al.}(2012)\citenamefont{Bodenstein,
  Dominguez, and Schilcher}}]{Bodenstein}
\bibinfo{author}{\bibfnamefont{S.}~\bibnamefont{Bodenstein}},
  \bibinfo{author}{\bibfnamefont{C.}~\bibnamefont{Dominguez}},
  \bibnamefont{and}
  \bibinfo{author}{\bibfnamefont{K.}~\bibnamefont{Schilcher}},
  \bibinfo{journal}{Phys.Rev.} \textbf{\bibinfo{volume}{D85}},
  \bibinfo{pages}{014029} (\bibinfo{year}{2012}), \eprint{1106.0427}.

\bibitem[{\citenamefont{Della~Morte and Juttner}(2010)}]{DellaMorte}
\bibinfo{author}{\bibfnamefont{M.}~\bibnamefont{Della~Morte}} \bibnamefont{and}
  \bibinfo{author}{\bibfnamefont{A.}~\bibnamefont{Juttner}},
  \bibinfo{journal}{JHEP} \textbf{\bibinfo{volume}{1011}}, \bibinfo{pages}{154}
  (\bibinfo{year}{2010}), \eprint{1009.3783}.

\bibitem[{pri()}]{priv}
\bibinfo{note}{{G. Hotzel, K. Jansen, M.Petschlies, private communication.}}

\bibitem[{\citenamefont{Burger et~al.}(2013)\citenamefont{Burger, Feng, Hotzel,
  Jansen, Petschlies et~al.}}]{Burger:2013jya}
\bibinfo{author}{\bibfnamefont{F.}~\bibnamefont{Burger}},
  \bibinfo{author}{\bibfnamefont{X.}~\bibnamefont{Feng}},
  \bibinfo{author}{\bibfnamefont{G.}~\bibnamefont{Hotzel}},
  \bibinfo{author}{\bibfnamefont{K.}~\bibnamefont{Jansen}},
  \bibinfo{author}{\bibfnamefont{M.}~\bibnamefont{Petschlies}},
  \bibnamefont{et~al.} (\bibinfo{year}{2013}), \eprint{1308.4327}.

\end{thebibliography}

\end{document}